\documentclass[]{pasj02} 
\usepackage[switch,mathlines]{lineno}
\usepackage{siunitx}
\usepackage{xcolor}

\jyear{2026}
\Received{}
\Accepted{}

\begin{document} 

\title{ A systematic study of CO/SiO absorption features in early-type galaxies using AKARI/IRC near-infrared spectra }

\author{
 Eiko \textsc{Kozaki},\altaffilmark{1}\altemailmark \email{kozaki\_e@u.phys.nagoya-u.ac.jp} 
 Takuma \textsc{Kokusho},\altaffilmark{1}
 Keiji \textsc{Nakayama},\altaffilmark{1}
 Shinki \textsc{Oyabu},\altaffilmark{2}
 Itsuka \textsc{Yachi},\altaffilmark{1}
 Keita \textsc{Yoshida},\altaffilmark{1}
 Shohei \textsc{Ono}\altaffilmark{1}
 and
 Hidehiro
 \textsc{Kaneda}\altaffilmark{1}
}
\altaffiltext{1}{Graduate School of Science, Nagoya University, Furo-cho, Chikusa-ku, Nagoya, Aichi 464-8602, Japan}
\altaffiltext{2}{Institute of Liberal Arts and Sciences, Tokushima University, 1-1 Minami-josanjima-cho, Tokushima-shi, Tokushima 770-8502, Japan}


\maketitle

\begin{abstract}
The origin of dust in early-type galaxies (ETGs) remains a long-standing question, with proposed sources being mass loss from evolved stars, galaxy mergers, or grain growth in the interstellar medium. To investigate the dominant source of dust in ETGs, we analyzed near-infrared spectra of 30 ETGs obtained with AKARI, focusing on the SiO and CO absorption features tracing the photospheres of old stellar populations. We also derived the dust mass using near- to far-infrared photometric data obtained by 2MASS, WISE, and AKARI. We find that the dust mass correlates with the summed equivalent widths of the SiO and CO absorption features.
This trend suggests that a significant fraction of dust in ETGs may originate from mass loss from evolved stars, consistent with an internal production scenario. The dust mass shows no anti-correlation with diffuse X-ray luminosities, suggesting that dust in ETGs is not strongly interacting with X-ray plasma. Moreover, polycyclic aromatic hydrocarbons (PAHs) are detected in the near-infrared spectra. We find that the PAH intensity shows no correlation with the equivalent widths of SiO and CO, but correlates with the luminosity of hot and warm dust components. This suggests that PAHs may be of external origin associated with galaxy merger remnants, heated by the activities of galactic nuclei. 
\end{abstract}


\section{Introduction}
Early-type galaxies (ETGs) are dominated by old stellar populations and generally considered to be quiescent systems with inactive star formation. In such systems, new dust production from young stars is expected to be inefficient. In addition, the interstellar medium (ISM) of ETGs is often filled with X-ray-emitting diffuse hot plasma (\cite{Forman1985}), which destroys dust grains through sputtering (\cite{Draine1979}; \cite{Tsai1995}). Therefore, ETGs were believed not to contain appreciable amounts of dust. \par
However, observations reveal that many ETGs contain considerable amounts of dust, typically an order of magnitude less than in spiral galaxies. In the far-infrared (far-IR), the first systematic detections of cold dust emission in ETGs were made by IRAS (e.g., \cite{Knapp1989}). Subsequent observations with ISO, Spitzer, and AKARI confirmed that cold dust is common in ETGs (e.g., Temi et al. \yearcite{Temi2003}, \yearcite{Temi2004}; \cite{Kaneda2011}). Herschel observations further revealed diverse and extended dust morphologies, suggesting a complex origin of dust in these systems (e.g., \cite{Smith2012}; \cite{di2013}). These studies find that the dust masses are higher than expected from the balance between production by stellar mass loss and destruction by sputtering in X-ray plasma (e.g., \cite{Goudfrooij1995}; Temi et al. \yearcite{Temi2004}, \yearcite{Temi2007}; \cite{Smith2012}; \cite{Kokusho2019}).\par
Despite the fact that polycyclic aromatic hydrocarbons (PAHs) are expected to be efficiently destroyed in X-ray plasma, PAH emission is also detected in ETGs. Mid-IR spectroscopy reveals that PAHs are present in a significant fraction of ETGs, often showing unusual band ratios compared to star-forming galaxies, suggesting differences in their physical conditions or excitation mechanisms (Kaneda et al. \yearcite{Kaneda2005}, \yearcite{Kaneda2008}; \cite{Rampazzo2013}). Spatially resolved studies further indicate that the PAH emission can be confined to dense or shielded regions, rather than following the stellar light distribution (\cite{Kaneda2011}). Statistical analyses using AKARI show that the PAH emission in ETGs is associated with systems containing molecular gas (\cite{Kokusho2017}).\par
The origin of the abundant dust in ETGs remains unclear, while the following two main hypotheses are proposed: internal and external origins. In the external origin scenario, ETGs are thought to acquire dust and cold gas through mergers with gas-rich, low-mass galaxies (e.g., \cite{Goudfrooij1994}; \cite{Davis2015}). This scenario is supported by several observational studies; optical imaging often reveals irregular dust morphologies, such as filaments and dust lanes, which are inconsistent with relaxed, internally produced dust components (\cite{Goudfrooij1994}). Interferometric CO observations show that the molecular gas in many ETGs has asymmetric and disturbed distributions, indicative of dynamically young systems (Young et al. \yearcite{Young2008}, \yearcite{Young2011}). Furthermore, the spatial and kinematic distributions of cold gas are frequently misaligned with those of the stellar component (Davis et al. \yearcite{Davis2011}, \yearcite{Davis2013}). \par
In contrast, the internal origin scenario predicts that dust is produced by stellar mass loss from evolved stars (\cite{Knapp1992}; \cite{Athey2002}; \cite{Temi2004}). It is also suggested that buoyant outflows from active galactic nuclei (AGN) transport dust to the outer regions of galaxies (\cite{Temi2007}). In the stellar mass loss scenario, evolved stars such as red giant branch and asymptotic giant branch (AGB) stars continuously inject gas and dust into the ISM through slow, dust-rich stellar winds. These winds offer a steady internal source of dust grains (\cite{Faber1976}; \cite{Mathews2003}). This mass loss from evolved stars is directly observed as emission in the 10--24\>µm wavelength range originating from the circumstellar regions of giant stars, and its surface brightness distribution closely follows that of the stellar light observed in the optical bands (\cite{Knapp1989}; \cite{Athey2002}). The growth of dust grains within the ISM is also suggested as a possible internal source of dust (\cite{Martini2013}; \cite{Hirashita2015}). \par
In the near-IR, the contribution of emission from late-type stars is dominant, and the effect of dust extinction is relatively minor. Therefore, a statistical analysis of near-IR spectra of ETGs provides a useful means to investigate stellar populations and atmospheric properties contributing to dust formation.
\citet{Cesetti2009} analyzed the near-IR (1.5--2.4\>µm) spectra of ETGs to study the absorption features as tracers of age and metallicity. These features are also considered to be useful tracers of the stellar mass loss.
However, no previous work statistically investigated the relationship between near-IR spectral properties and dust formation in ETGs. In this study, we combine near-IR spectroscopy obtained with the Infrared Camera (IRC; \cite{Onaka2007}) onboard AKARI with mid- and  far-IR photometry to investigate the origin of dust in ETGs.

\section{Observations and data analysis}\label{sec:2}
 
\subsection{Data and sample}\label{ssec:21}
The AKARI/IRC near-IR spectra cover the wavelength range of 2.5--5.0 µm with a spectral resolution of $R\sim$120, and achieve typical sensitivities of a few mJy (e.g., \cite{Onaka2007}; \cite{Ohyama2007}). The near-IR spectra of the ETGs analyzed in this study exhibit broad absorption features at wavelengths longer than 4.0\>µm. These features are thought to originate from the ro-vibrational transitions of SiO ($\Delta v = 2$, 4.3\>µm) and CO ($\Delta v = 1$, 4.66\>µm) present in the photospheres of late-type stars (\cite{Heras2002}; \cite{Gautschy2004}). An additional absorption feature is seen at around 2.5\>µm, which is attributed to the ro-vibrational transition of CO ($\Delta v = 2$) (\cite{Heras2002}; \cite{Gautschy2004}).
In this study, we focus on ETGs observed with AKARI/IRC that are classified as quiescent based on the slope of their near-IR continuum. We further require that near-IR spectra show the SiO and CO absorption features tracing stellar mass loss and have sufficient spectral quality around 3.3\>µm to discuss the PAH emission. Among the 804 galaxies for which near-IR spectra were obtained with AKARI/IRC, 52 objects belong to ETGs that were detected by IRAS (\cite{Knapp1989}). From these, we selected 30 ETGs that satisfy the above criteria, as listed in table 1. The sample ETGs have stellar masses of $10^{10}$--$10^{11}\,M_\odot$, consisting of 22 elliptical (E) and 8 lenticular (S0) galaxies. In terms of the environment, two galaxies are classified as field galaxies, while the remaining 28 are associated with group or cluster environments. Since the selection is based on the IRAS detections, the sample ETGs are biased toward relatively dust-rich ETGs. \par

\begin{table}[h]
  \tbl{The sample of 30 ETGs. 
  }{%
  \begin{tabular}{ccccc}
      \hline
      Name & Type & Distance (Mpc)\footnotemark[$*$] & SiO EW (nm) & CO EW (nm)\\ 
      \hline
      IC\,3370 & E & 26.8 & 78\,$\pm$\,2 & 78\,$\pm$\,3 \\
      NGC\,0315 & E & 51.0 & 69\,$\pm$\,5 & 102\,$\pm$\,6 \\
      NGC\,1052 & E & 19.2 & 68\,$\pm$\,2 & 46\,$\pm$\,4 \\
      NGC\,1291 & S0 & 9.08 & 92\,$\pm$\, 3 & 119\,$\pm$\,4 \\
      NGC\,1316 & S0 & 20.8 & 79\,$\pm$\,2 & 119\,$\pm$\,3 \\
      NGC\,1549 & E & 18.2 & ...\footnotemark[$\dagger$] & ...\footnotemark[$\dagger$] \\
      NGC\,2768 & E & 22.2 & 76\,$\pm$\,3 & 40\,$\pm$\,5 \\
      NGC\,3377 & E & 10.6 & 53\,$\pm$\,13 & 0\,$\pm$\,4 \\
      NGC\,3894 & E & 48.2 & 84\,$\pm$\,10 & 96\,$\pm$\,9 \\
      NGC\,3904 & E & 28.3 & 67\,$\pm$\,6 & 87\,$\pm$\,7 \\
      NGC\,3962 & E & 31.2 & 102\,$\pm$\,5 & 56\,$\pm$\,7 \\
      NGC\,3998 & S0 & 14.2 & 69\,$\pm$\,6 & 82\,$\pm$\,8 \\
      NGC\,4125 & E & 24.0 & 76\,$\pm$\,3 & 110\,$\pm$\,5 \\
      NGC\,4203 & S0 & 15.3 & 132\,$\pm$\,11 & 80\,$\pm$\,13 \\
      NGC\,4261 & E & 32.4 & 82\,$\pm$\,4 & 109\,$\pm$\,5 \\
      NGC\,4278 & E & 15.4 & 91\,$\pm$\,5 & 85\,$\pm$\,6\\
      NGC\,4406 & E & 17.1 & 101\,$\pm$\,7 & 101\,$\pm$\,9 \\
      NGC\,4486 & E & 16.1 & 80\,$\pm$\,4 & 97\,$\pm$\,5 \\
      NGC\,4552 & E & 15.9 & 81\,$\pm$\,5 & 126\,$\pm$\,6 \\
      NGC\,4636 & E & 15.1 & 87\,$\pm$\,5 & 110\,$\pm$\,7 \\
      NGC\,4696 & S0 & 35.5 & 48\,$\pm$\, 4 & 73\,$\pm$\,6 \\
      NGC\,4697 & E & 11.8 & 75\,$\pm$\,2 & 102\,$\pm$\,3 \\
      NGC\,5018 & E & 33.9 & 77\,$\pm$\, 4 & 85\,$\pm$\,4 \\
      NGC\,5044 & E & 32.3 & 63\,$\pm$\, 3 & 104\,$\pm$\,5 \\
      NGC\,5353 & S0 & 27.8 & 122\,$\pm$\,12 & 82\,$\pm$\,13 \\
      NGC\,5363 & S0 & 16.6 & 144\,$\pm$\,6 & 134\,$\pm$\,7 \\
      NGC\,5813 & E & 32.2 & 28\,$\pm$\,6 & 52\,$\pm$\,7 \\
      NGC\,6482 & E & 57.6 & 108\,$\pm$\,6 & 131\,$\pm$\,7 \\
      NGC\,6703 & S0 & 27.2 & 88\,$\pm$\,8 & 124\,$\pm$\,7 \\
      NGC\,7052 & E & 46.4 & 109\,$\pm$\,14 & 167\,$\pm$\,18 \\
      \hline
    \end{tabular}}
\begin{tabnote}
\footnotemark[$*$] Distances are adopted from the NASA/IPAC Extragalactic Database, using redshift-independent measurements.  \\ 
\footnotemark[$\dagger$] The equivalent width cannot be derived due to contamination by artifacts in the spectrum.\\ 
 
\end{tabnote}
\end{table}

The AKARI satellite was equipped with two scientific instruments: IRC and Far-infrared surveyor (\cite{Kawada2007}). Launched on 2006 February 22, AKARI operated for about 550 days in the "cold phases", during which the instruments were cooled down to approximately 6\>K using liquid helium (LHe) and mechanical coolers, and all-sky surveys were conducted. 
After the depletion of LHe, only the mechanical coolers were used for cooling, resulting in an increase in the overall instrument temperature to 40--50\>K. This period, referred to as the "warm phase", was dedicated solely to pointed observations with IRC.\par
The IRC data reduction is based on the official pipeline provided by the AKARI team. However, during the warm phase, particularly when the detector temperature exceeds 45\>K, the dark current becomes highly sensitive to small temperature variations, causing the standard dark-frame subtraction to fail for warm-phase data. To address this issue, a new dark subtraction method has been adopted for IRC, which utilizes the temperature dependence of the dark current for each pixel, derived from laboratory tests using the IRC engineering-model detector by \citet{Mori2011}.
They determined the effective detector temperature by minimizing  the standard deviation of the dark-subtracted data to precisely perform the dark-frame subtraction.
In this study, we use near-IR spectra processed with this method to investigate the properties of stellar atmospheres in ETGs. To estimate the dust properties of the sample ETGs, we also use the photometric data obtained by 2MASS, WISE, and AKARI.

\subsection{Near-IR spectral fitting}\label{ssec:22}
\begin{figure*}[ht]
 \begin{center}
  \includegraphics[width=16cm]{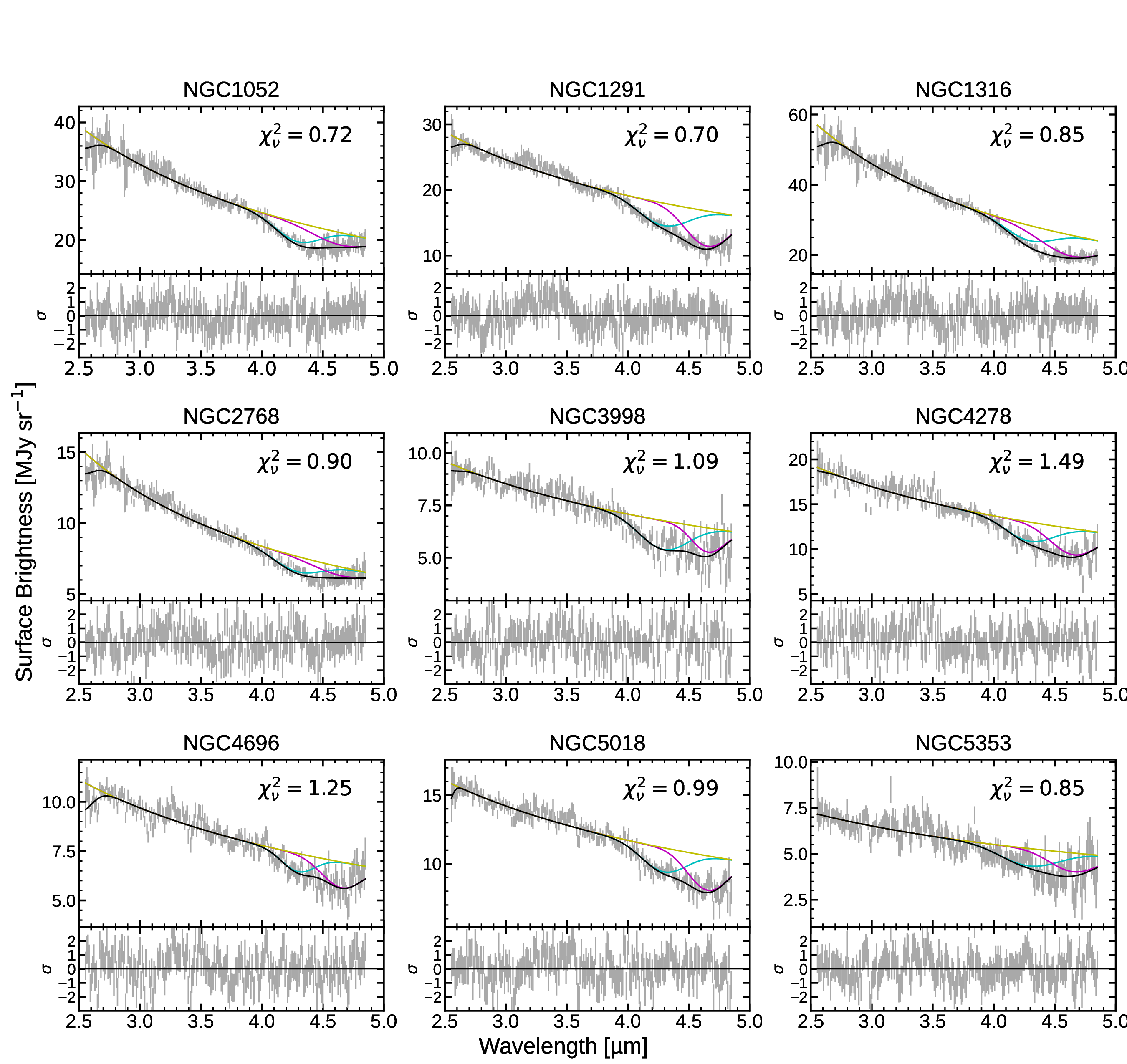} 
 \end{center}
\caption{Examples of the spectral fitting results for the AKARI/IRC 2.5--5.0 µm spectra. The spectral model (black) consists of the continuum component (yellow), SiO ($\Delta v=2$) absorption feature (cyan), CO ($\Delta v = 1$) absorption feature (magenta) and CO ($\Delta v=2$) absorption feature. The bottom panels show the residuals obtained by subtracting the best-fit model from observational data, normalized by the uncertainties; the vertical axis is expressed in units of $\sigma$.
 {Alt text: Nine panels of examples of the spectral fitting.} 
}\label{fig:1}
\end{figure*}
In this study, in order to estimate the abundances of SiO and CO molecules in the stellar atmospheres, we fit the spectra using the following function:
\begin{equation}
  F_\nu =  F_\mathrm{continuum} e^{-\tau_{\nu}}, 
  \label{eq1}
\end{equation}
where $F_\mathrm{continuum}$ and $\tau_{\nu}$ represent the continuum emission and the optical depth due to SiO and CO absorption, respectively. The continuum component is assumed to arise from stellar emission and hot dust, and is modeled using a power-law function defined as $F_\mathrm{continuum} \propto \lambda^{\Gamma}$. For the absorption features of SiO and CO, we adopt Gaussian profiles, which are defined as
\begin{eqnarray}
\tau_{\nu}&=&\tau_{\mathrm{SiO}\left(\Delta v=2\right)} + \tau_{\mathrm{CO} \left(\Delta v=1\right)} + \tau_{\mathrm{CO} \left(\Delta v=2\right)} \nonumber \\
&=&\sum_{i=1}^3 \alpha_i \; \mathrm{exp}\left(-4\;\mathrm{ln}\;2 \;\frac{\left(\lambda - \lambda_{r, i}\right)^2}{\gamma^2}\right), 
\end{eqnarray}
where $\alpha$, $\gamma$, and $\lambda_{r}$ are the peak, the full width at half maximum, and the central wavelength of the Gaussian profile, respectively.\par

The fitting is performed using a non-linear least-squares method with MPFIT (\cite{Markwardt2009}). The fitting procedure consists of the following three steps: first, the continuum component alone is fitted with a power-law model over the wavelength ranges of 2.8--3.2 and 3.4--4.0\>µm. 
Second, the SiO and CO absorption features are added to the model with the power-law component fixed at the best-fit model in the first step. The central wavelengths of the absorption features are fixed at 4.3, 4.66, and 2.5\>µm, while their widths and amplitudes are allowed to vary. 
Finally, the continuum parameters and the amplitudes of the absorption features are refitted simultaneously, with the widths of the absorption features fixed at the best-fit model in the second step.
From these fits, we derive the equivalent widths (EWs) of the SiO and CO absorption features. Since the CO ($\Delta v=2$) absorption feature at 2.5\>µm is only partially covered in our data, the fitting result of this feature is not used in this study. Figure 1 shows examples of the final fitting results. The residual spectra exhibit a certain level of correlated noise, which may be related to instrumental effects, such as imperfect dark subtraction. \par
By stacking the residuals between the fitted model and the data for all galaxies, we find that a systematic excess appears in the 3.2--3.6\>µm range. Interpreting this component as PAH band emission, we fit the stacked residual spectrum using the same method as \citet{Kondo2024}, incorporating the 3.3\>µm aromatic hydrocarbon feature and the 3.4--3.6\>µm aliphatic hydrocarbon features, as shown in figure 2.
For the aromatic feature, they adopt the Drude profile.
The aliphatic features consist of four components at central wavelengths of 3.41, 3.46, 3.51, and 3.56\>µm. The first sub-component is enough to be resolved with the spectral resolution of AKARI/IRC and is therefore modeled with a Lorentzian profile, while the remaining three are unresolved and thus modeled with Gaussian profiles.
We fit the stacked residual spectrum by treating the central wavelengths, widths, and amplitudes of the PAH components as free parameters. The central wavelengths and widths obtained from this fit are then fixed and incorporated into the model of equation 1 as follows:
\begin{eqnarray}
 F_\nu =  \left(F_\mathrm{continuum} +  F_\mathrm{PAH3.3}\right)\;e^{-\tau_{\nu}}   . 
\end{eqnarray}
Finally, we fit each individual spectrum by allowing the continuum component and the amplitudes of the absorption and PAH components to vary freely. From these fits, we derive the intensity of the PAH feature $I_\mathrm{PAH}$.
\begin{figure}
 \begin{center}
  \includegraphics[width=8cm]{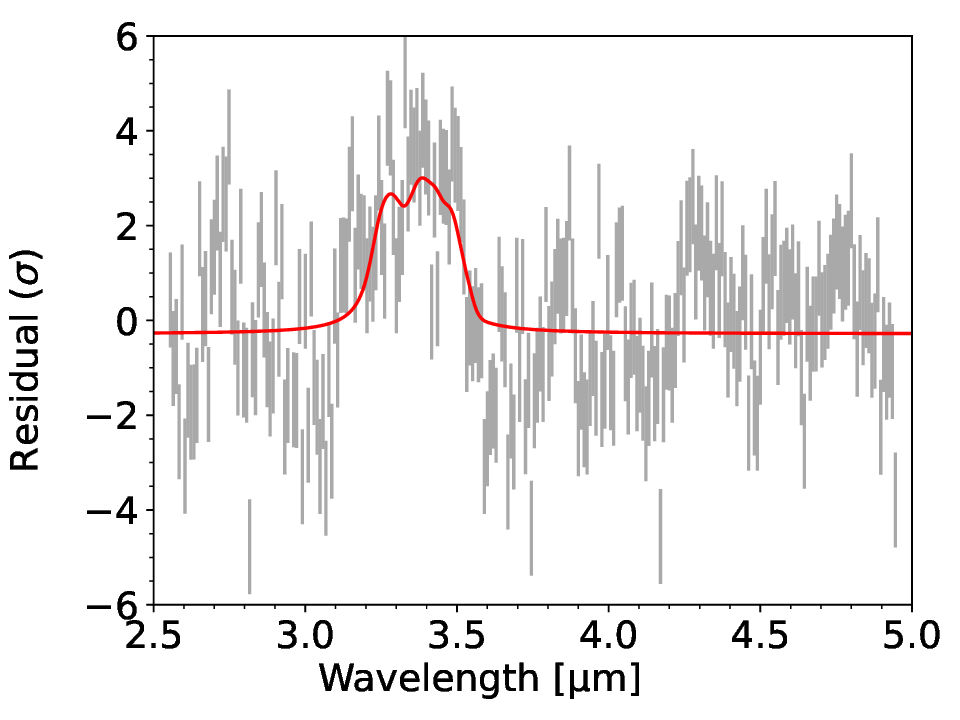} 
 \end{center}
 \caption{Stacked residual spectrum for the 29 sample ETGs, excluding one galaxy affected by artifacts.  Gray bars show the stacked residuals, defined as the difference between the observed spectra and fitted models normalized by the 1$\sigma$ uncertainty of the stacked spectrum. Red line shows the fitted PAH model (see text for details). 
 {Alt text: One graph.} }
\label{fig:2}
\end{figure}

\subsection{Spectral energy distribution fitting}\label{ssec:23}
To investigate the properties of cold dust in the sample ETGs, we create the spectral energy distributions (SEDs) using the near- to far-IR photometry data obtained with AKARI, WISE, and 2MASS. For the 2MASS data, we retrieved the $K_\mathrm{s}$-band flux from the 2MASS Extended Source Catalog (\cite{Jarrett2000}). In addition, using the effective radii in the near-IR bands ($\mathrm{j}\_\mathrm{r}\_\mathrm{eff}$, $\mathrm{h}\_\mathrm{r}\_\mathrm{eff}$, $\mathrm{k}\_\mathrm{r}\_\mathrm{eff}$) and the major-to-minor axis ratio  in the $K_\mathrm{s}$ band ($\mathrm{k}\_\mathrm{ba}$), we calculate the effective radius in the optical $B$ band as $R_\mathrm{e} = 1.7 \times \mathrm{MEDIAN}(\mathrm{j}\_\mathrm{r}\_\mathrm{eff}, \mathrm{h}\_\mathrm{r}\_\mathrm{eff}, \mathrm{k}\_\mathrm{r}\_\mathrm{eff})\sqrt{\mathrm{k}\_\mathrm{ba}}$ for the aperture photometry of the AKARI data (\cite{Cappellari2011}). For the WISE data, we obtained the fluxes in the four bands with the effective wavelengths of 3.4, 4.6, 12, and 22\>µm from the AllWISE catalog (\cite{Cutri2013}).\par
For the AKARI data, we use the mid-IR (9 and 18\>µm) and far-IR (65, 90, and 140\>µm) bands, and derive the flux densities as described below. First, we perform aperture photometry using a circular aperture defined as $R_\mathrm{aper}=\sqrt{(2R_e)^2+(1.5D_\mathrm{PSF})^2}$, where $D_\mathrm{PSF}$ denotes the full width at half maximum of the point spread function at each band. To estimate the background level, we define the background region as an annulus between $1.5R_\mathrm{aper}$ and $2.5R_\mathrm{aper}$. Because the background region can be contaminated by other sources or artifacts, we apply 3\,$\sigma$ clipping when estimating the background level. The systematic uncertainties in the absolute flux calibration are 10\% for the mid-IR bands, and 20\%, 20\%, and 50\% for the far-IR bands at 65, 90, and 140\>µm, respectively (Takita et al. 2015).\par
To estimate the dust mass of the sample ETGs, we fit each SED with a model composed of emission from stars and three dust components with different temperatures (hot, warm, and cold). The stellar continuum is modeled with a power-law with the index fixed at $-2$. In addition, ETGs are known to exhibit a silicate feature around 10 µm (\cite{Bressan2006}), and we include this feature in the model using a Gaussian profile. The amplitude relative to the stellar continuum and the width of the Gaussian are determined based on the template for quiescent elliptical galaxies presented in \citet{Kaneda2008}. Each of the three dust components is modeled using a modified blackbody with the emissivity power-law index of two.
For some ETGs, flux densities are not available at all wavelengths due to background noise and other effects, making it difficult to determine both the temperature and the amplitude of the dust emission simultaneously. Therefore, we adopt the following fitting procedure: first, we stack the SEDs of galaxies for which complete wavelength coverage is available. Second, we fit the stacked SED by treating the temperatures and amplitudes of the three dust components as free parameters to determine the dust temperatures for the sample ETGs. Finally, we fit the individual SEDs by fixing these parameters and allowing only the amplitudes to vary. Figure 3 shows examples of the SED fitting results. \par
\begin{figure*}
 \begin{center}
  \includegraphics[width=16cm]{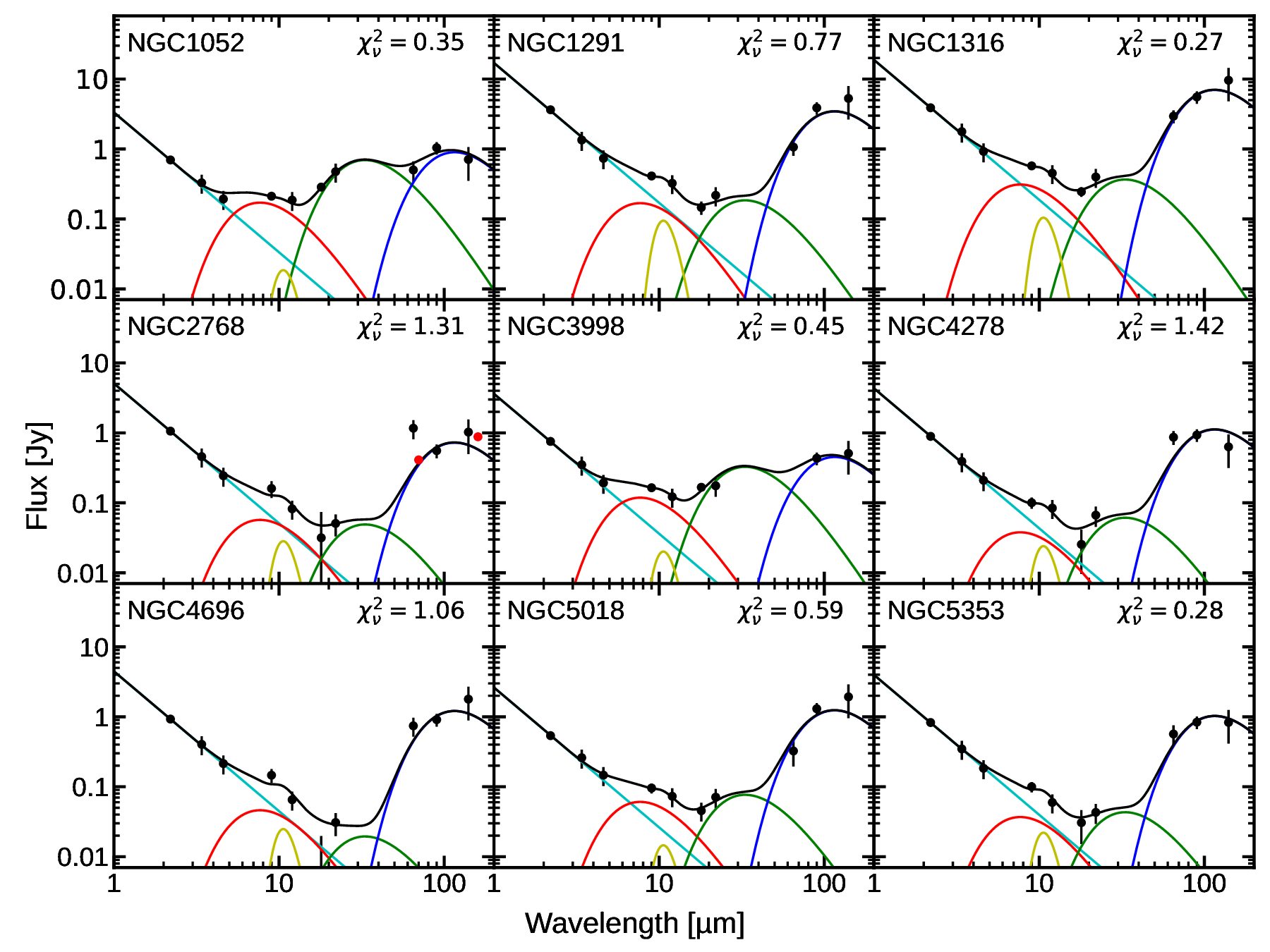} 
 \end{center}
\caption{Examples of the SED fitting of the same sample ETGs as shown in figure 1. The SED model (black) consists of stellar continuum (cyan), silicate feature (yellow), hot (red), warm (green) and cold dust (blue). For NGC\,2768, additional Herschel data points at 70 and 160\,µm are overplotted as red symbols.
 {Alt text: Nine panels of the SED fitting.} 
}\label{fig:3}
\end{figure*}
The amplitudes derived from these fits are used to compute the dust mass for each dust component. The flux density of dust emission is expressed as 
\begin{equation}
 F_\nu = \frac{\kappa_\nu \ M_\mathrm{dust}\  B_\nu(T_\mathrm{dust})}{D^2}, 
  \label{eq1}
\end{equation}
and the dust mass is derived through SED fitting as
\begin{equation}
  M_\mathrm{dust} = \frac{F_\nu \ D^2}{\kappa_\nu \ B_\nu(T_\mathrm{dust})}, 
  \label{eq1}
\end{equation}
where $T_\mathrm{dust}$ is the dust temperature, $B_\nu(T_\mathrm{dust})$ is the Planck function, $D$ is the distance to the galaxy, and the dust mass absorption coefficient $\kappa_\nu$ is adopted from \citet{Draine2003}.\par
Some of the sample ETGs have archival Herschel far-IR data, which can be added to their SEDs. For NGC\,2768 and NGC\,4125, for instance, their AKARI-based SED fits appear relatively uncertain in the far-IR, and thus we perform aperture photometry on the Herschel data to derive their flux densities at 70, 100, and 160\,µm. We confirm that the cold dust masses derived with and without the Herschel data agree within 2$\sigma$ uncertainties, suggesting that the inclusion of the Herschel data may not significantly affect our results. Thus, in order to maintain a homogeneous analysis, we adopt the AKARI far-IR all-sky survey data alone for the full sample.

\section{Results}\label{sec:3}
\subsection{Relationship between SiO and CO absorption features and dust mass}
From the near-IR spectra fitting, we successfully measured the EWs of the SiO and CO absorption features for 29 of the 30 sample ETGs. For one galaxy, the spectrum around the SiO and CO absorption features is significantly affected by artifacts, preventing reliable EW measurements; this galaxy is therefore excluded from the following results. The measured EWs of the sample ETGs are listed in table 1. These absorption features are thought to trace molecular gas present in the photospheres of AGB stars.
\begin{figure*}
 \begin{center}
  \includegraphics[width=16cm]{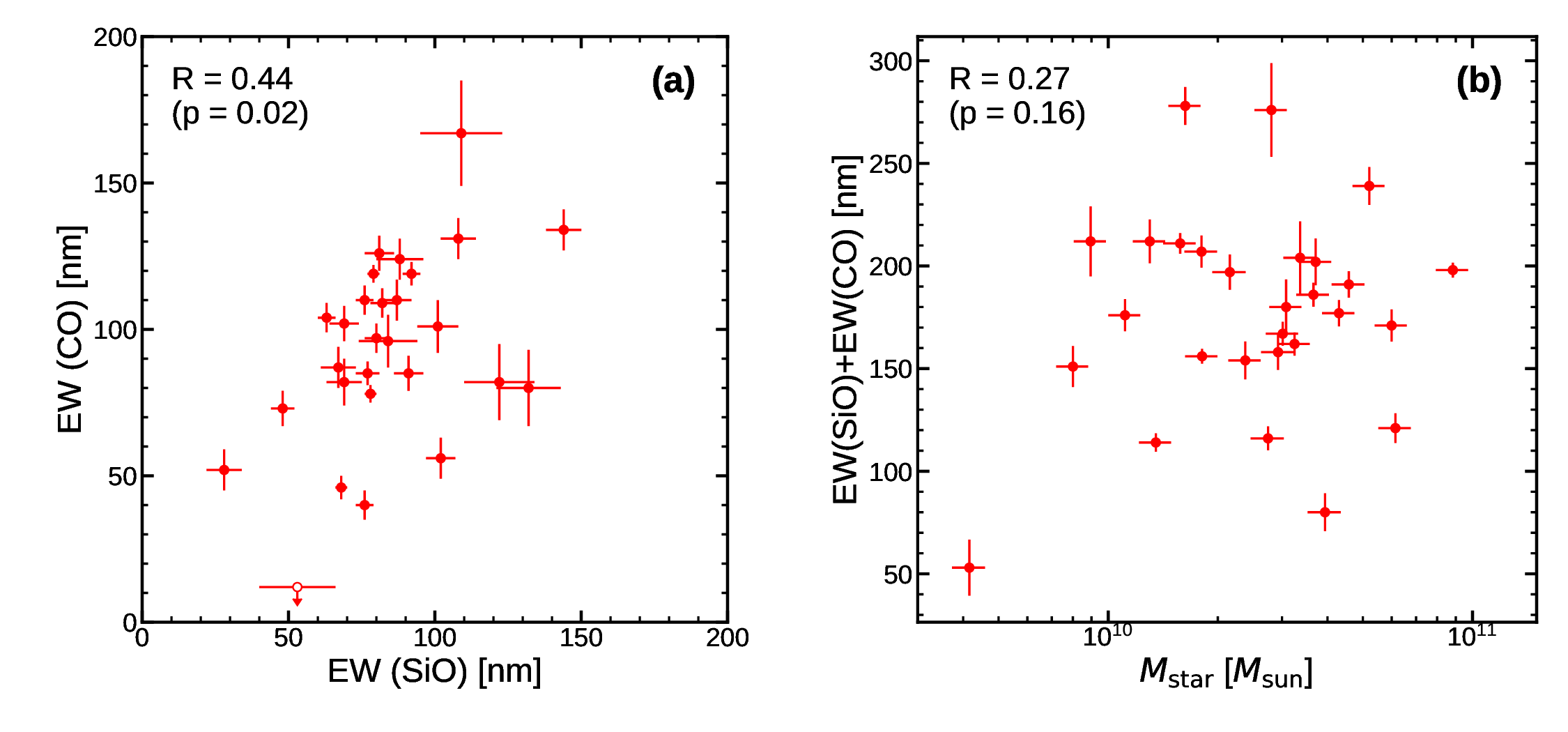} 
 \end{center}
\caption{(a) Scatter plot of EW(CO) vs EW(SiO). The upper limit corresponds to 3$\sigma$. The numbers shown in the upper-left corner indicate the Pearson correlation coefficient (R) and the corresponding p-value, calculated using the same scale as in the plot. This notation is used in all the subsequent figures. (b) Scatter plot of EW(SiO)+EW(CO) vs $M_\mathrm{star}$.
 {Alt text: Two scatter plots.} 
}\label{fig:4}
\end{figure*}
Figure 4a shows a relationship between EW(CO) and EW(SiO), suggesting a correlation between the two EWs. 
In O-rich AGB stars, after carbon is bound to oxygen to form CO, the remaining oxygen combines with silicon to produce SiO (e.g., Tsuji \yearcite{Tsuji1973}, \yearcite{Tsuji1978}). Therefore, O-rich AGB stars are expected to exhibit both SiO and CO absorption features (e.g., \cite{Origlia1993}; \cite{Rayner2009}), and the correlation shown in figure 4a suggests that the sample ETGs are dominated by the O-rich stellar populations. This result is consistent with model expectations that environments dominated by low-mass, metal-rich stellar populations, such as ETGs, preferentially harbor O-rich AGB stars (e.g., \cite{Mouhcine2003}; \cite{Marigo2007}; \cite{Karakas2010}). Figure 4b shows the summed EWs of the SiO and CO absorption features as a function of the galaxy stellar mass, which is estimated from the 2MASS $K_\mathrm{s}$-band luminosities using the typical $K$-band mass-to-light ratio $M_\mathrm{star}/L_K=1.21 \,M_\odot/L_\odot$ (\cite{Cappellari2013}). As seen in the figure, there is no systematic trend with $M_\mathrm{star}$, indicating that the observed changes in the EW are not primarily driven by the luminosities related to the stellar masses, but indeed by galaxy-to-galaxy variations in the contribution from the mass loss of O-rich AGB stars in our sample.\par
From the SED fitting covering the near- to far-IR wavelengths, we derive the physical properties of the three dust components with different temperatures for 29 of the 30 sample ETGs. One galaxy is excluded from the SED analysis because stripe-like artifacts in the AKARI far-IR images prevent reliable aperture photometry.
\begin{figure}[t]
 \begin{center}
  \includegraphics[width=8cm]{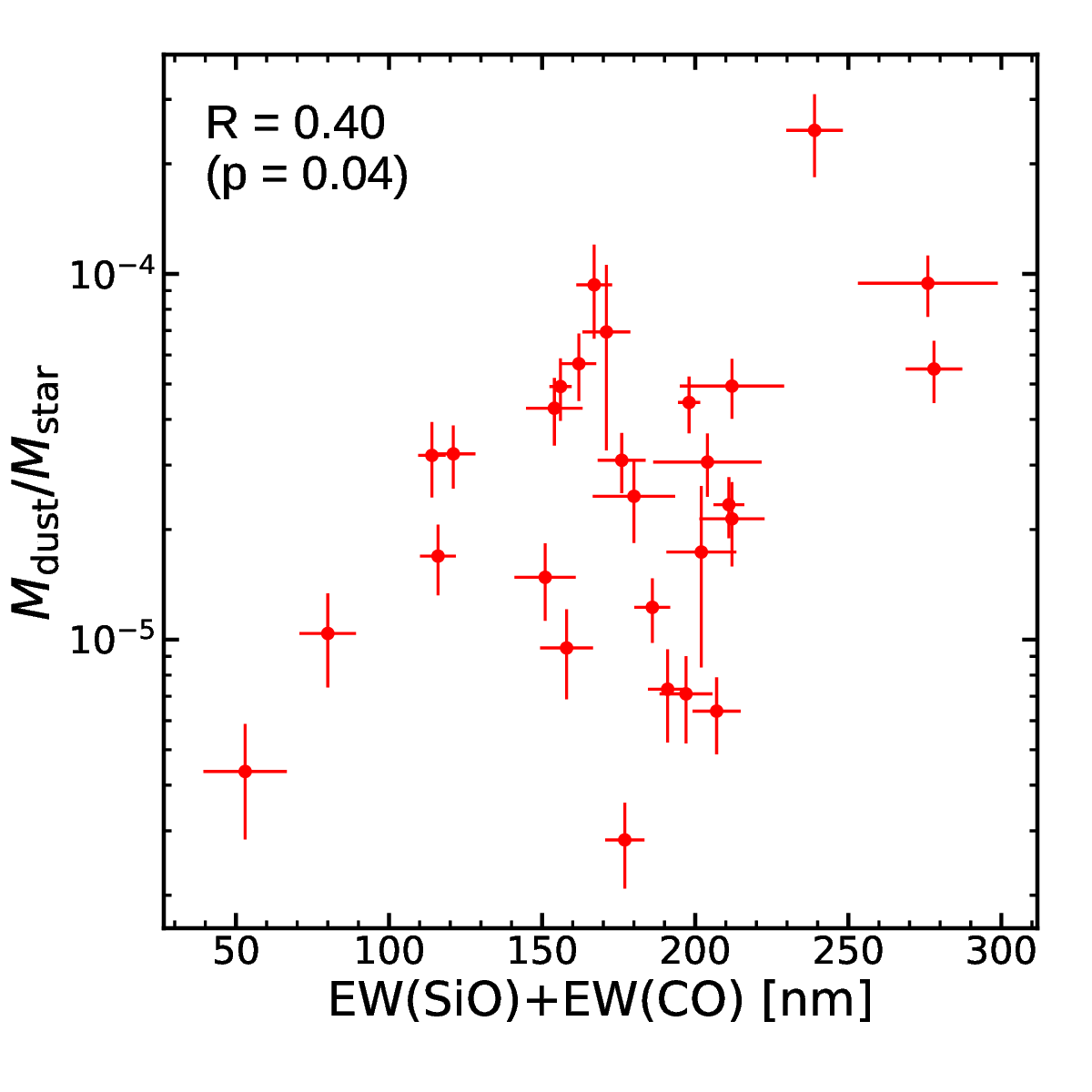} 
 \end{center}
\caption{ Scatter plot of $M_\mathrm{dust}$/$M_\mathrm{star}$ vs EW(SiO)+EW(CO).
 {Alt text: One scatter plot.} 
}\label{fig:5}
\end{figure}
Figure 5 shows the relationship between the dust mass normalized by the stellar mass, $M_\mathrm{dust}/M_\mathrm{star}$, and the summed EWs of the SiO and CO absorption features. 
The dust mass is obtained by summing over the three components.  
Figure 5 suggests a correlation between $M_\mathrm{dust}/M_\mathrm{star}$ and the sum of the EWs, although the correlation is relatively weak (correlation coefficient $\sim$0.4 with a $p$-value of $\sim$0.04). Since, as shown in figure 4b, $M_\mathrm{star}$ is not essential to cause the changes in the EWs, the result of figure 5 suggests that the stellar mass loss may be an important dust source in the sample ETGs. In addition, there is a scatter of about 1--1.5 dex at a given EW, which may indicate non-negligible contributions from additional dust sources.

\subsection{Relationship between SiO and CO absorption features and PAH emission}
We detect the 3.3\>µm PAH emission feature in 22 out of the 30 sample ETGs. Here, we consider PAHs to be detected when inclusion of the PAH component significantly improves the spectral fit according to an F-test at the 90\% confidence level. 
\begin{figure}
 \begin{center}
  \includegraphics[width=8cm]{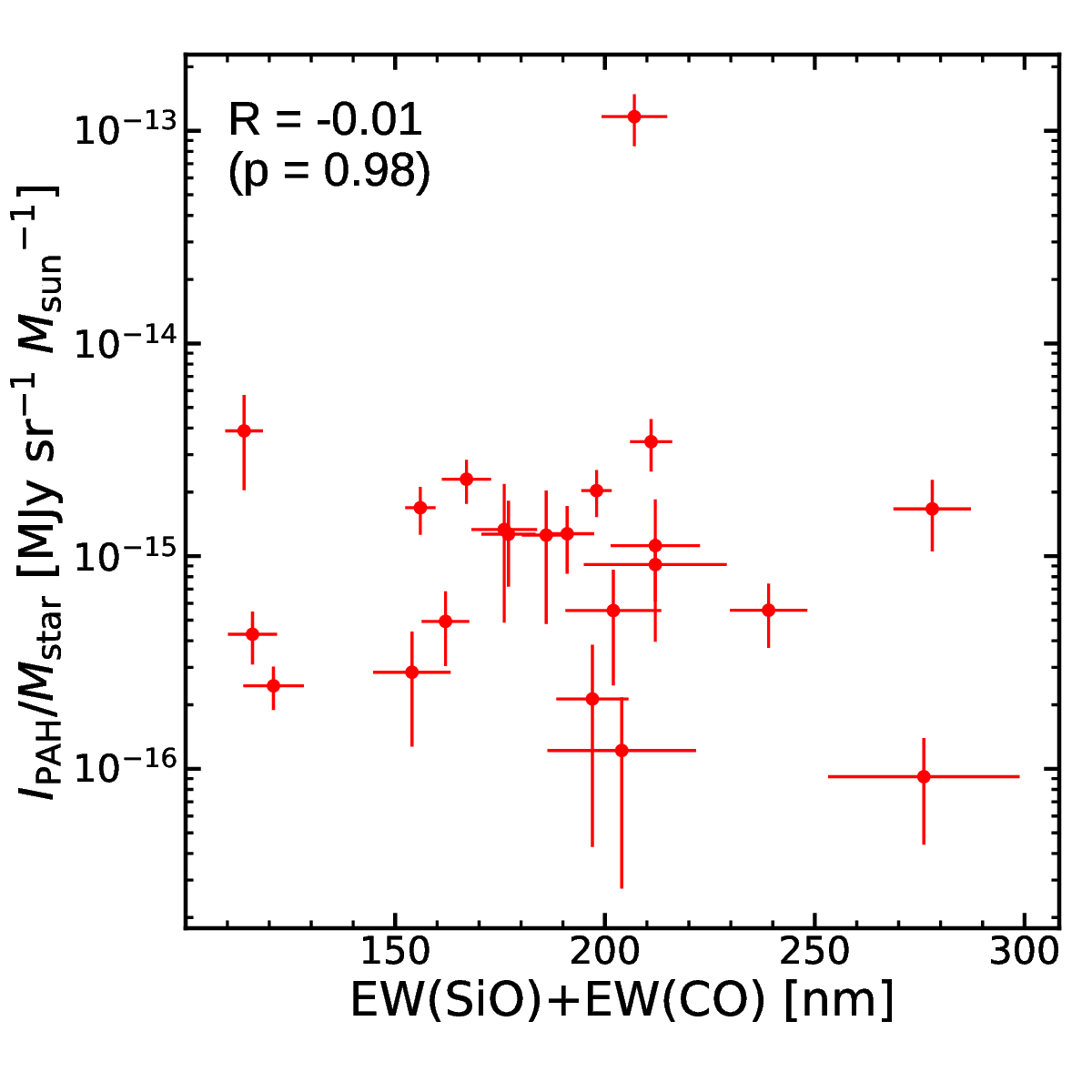} 
 \end{center}
\caption{Scatter plot of $I_\mathrm{PAH}/M_\mathrm{star}$ vs EW(SiO)+EW(CO) for the sample ETGs detected in the PAH emission. 
{Alt text: One scatter plot.} 
}\label{fig:6}
\end{figure}
Figure 6 shows the relationship between the PAH intensity normalized by $M_\mathrm{star}$ and the summed EWs of the SiO and CO absorption features. As shown in this figure, no correlation is found between the PAH intensity and the molecular absorption features, suggesting that PAHs are not directly associated with the stellar mass loss. This result further implies that the PAH emission traces a population of dust different from that of the cold dust detected in the far-IR.

\section{Discussion}\label{sec:4}

\subsection{The origin of the dust in ETGs}
The results obtained in this study suggest that 
the mass loss from evolved stars is an important source of dust in the sample ETGs. The correlation between the dust mass and the summed EWs of SiO and CO absorption features (figure 5) suggests that the dust is associated with evolved stellar populations such as AGB stars. Since these molecular absorptions trace the atmospheres of O-rich AGB stars (figure 4a), their EWs are considered to reflect the contribution from stellar mass loss. \par
\begin{figure}
 \begin{center}
 \includegraphics[width=8cm]{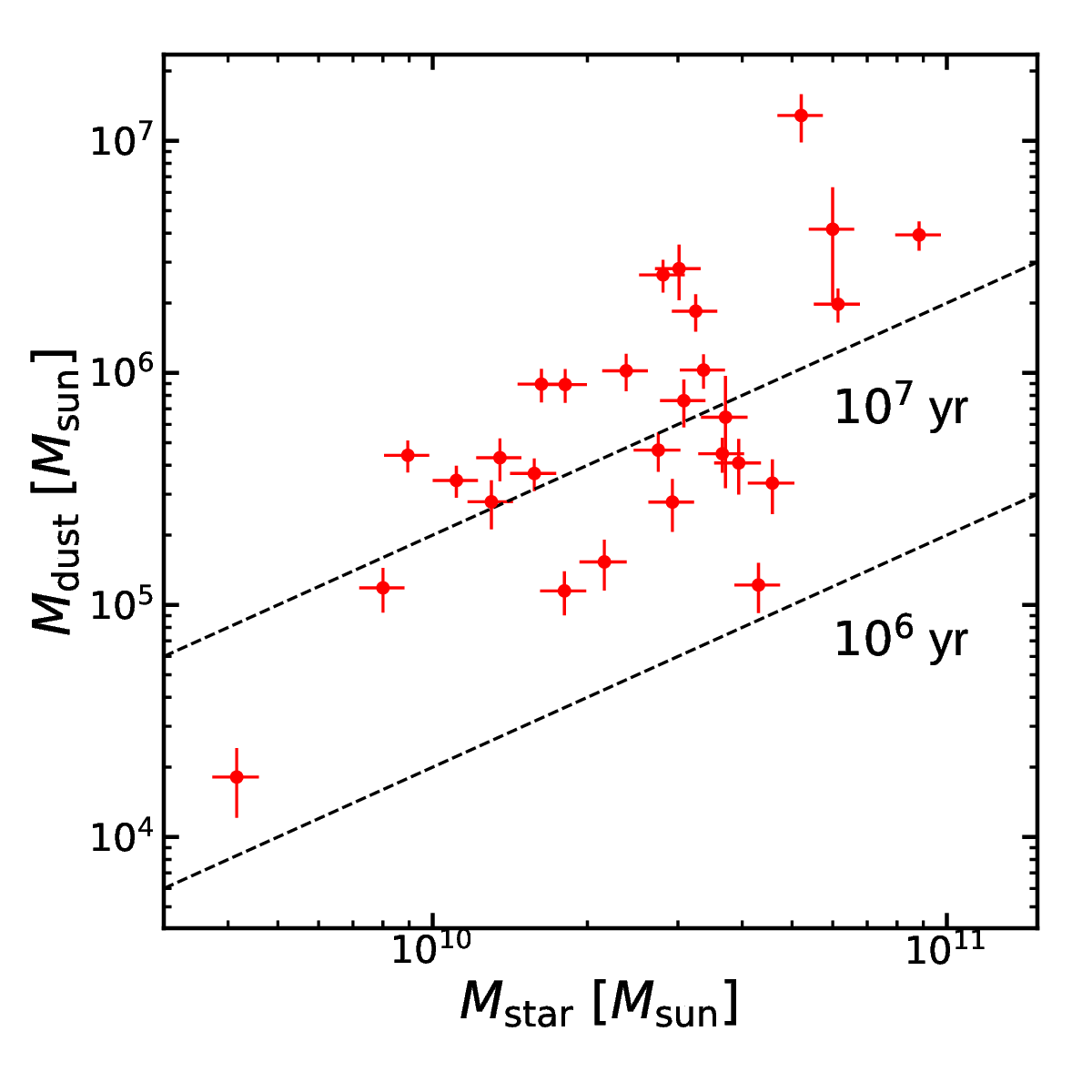} 
 \end{center}
\caption{ Scatter plot of $M_\mathrm{dust}$ vs $M_\mathrm{star}$. The two dotted lines indicate dust masses expected from the balance between the production by stellar mass loss and the destruction by sputtering in X-ray plasma (see text for details).
 {Alt text: One scatter plot.} 
}\label{fig:7}
\end{figure}
Figure 7 shows the relation between $M_\mathrm{dust}$ and $M_\mathrm{star}$ for the sample ETGs. 
The dotted line in the figure represents the dust mass expected when the dust supply from stellar mass loss is balanced by the sputtering destruction in X-ray plasma. The dust supply rate, $\dot{M}$, is given by \citet{Knapp1992} as follows:
\begin{equation}
  \dot{M} = 2.1\times10^{-12}\left(\frac{L_\mathrm{K}}{L_\odot}\right) [M_\odot \ \mathrm{yr^{-1}}]. 
  \label{eq}
\end{equation}
The timescale for the dust destruction by sputtering in X-ray plasma is given by Draine and Salpeter (\yearcite{Draine1979}) as follows:
\begin{equation}
  \tau = 2\times10^5 \left(\frac{\mathrm{cm^{-3}}}{n_\mathrm{H}}\right) \left(\frac{a}{\SI{0.1}{\>\micro \meter}}\right) [\mathrm{yr}], 
  \label{eq}
\end{equation}
where $n_\mathrm{H}$ denotes the proton density of the plasma, and $a$ represents the dust grain radius. Assuming a typical dust grain size of $a=0.1$\>µm, we derive the dust mass $M_\mathrm{dust}=\dot{M}\tau$ for the plasma proton densities of $n_\mathrm{H}=10^{-1}$ and $10^{-2}$\>cm$^{-3}$ (e.g. \cite{Trinchieri1986}; \cite{Irwin1996}) as shown in figure 7. 
X-ray plasma is expected to efficiently destroy dust grains through sputtering. 
However, more than half of the sample ETGs exhibit dust masses exceeding those expected from the balance between stellar mass loss and sputtering destruction in hot plasma (figure 7),
suggesting that the process of dust destruction by X-ray plasma is not that efficient. 
Furthermore, as shown in figure 8, no significant anti-correlation is found between $M_\mathrm{dust}/M_\mathrm{star}$ and the X-ray luminosity normalized by stellar mass, $L_X/M_\mathrm{star}$, for the 18 galaxies with available Chandra diffuse X-ray measurements (\cite{Boroson2011}; \cite{Kim2015}; \cite{Su2015}; \cite{Goulding2016}). If dust were strongly interacting with the X-ray plasma, galaxies with higher X-ray luminosities would be expected to contain less dust. The absence of such a trend implies that the X-ray plasma does not fill the whole interstellar space of the ETGs, or alternatively that the dust resides in a cold gas phase shielded from the X-ray plasma. \par
In addition, since the SiO and CO absorption features used in this study are tracers of O-rich AGB stellar atmospheres, the stellar mass loss discussed here is expected to be predominantly producing silicate dust in chemical composition. Therefore, the results based on figures 4a and 5 suggest that the significant fraction of the detected dust may be in the form of the cold silicate dust originating from the mass loss of evolved stars.
\begin{figure}
 \begin{center}
  \includegraphics[width=8cm]{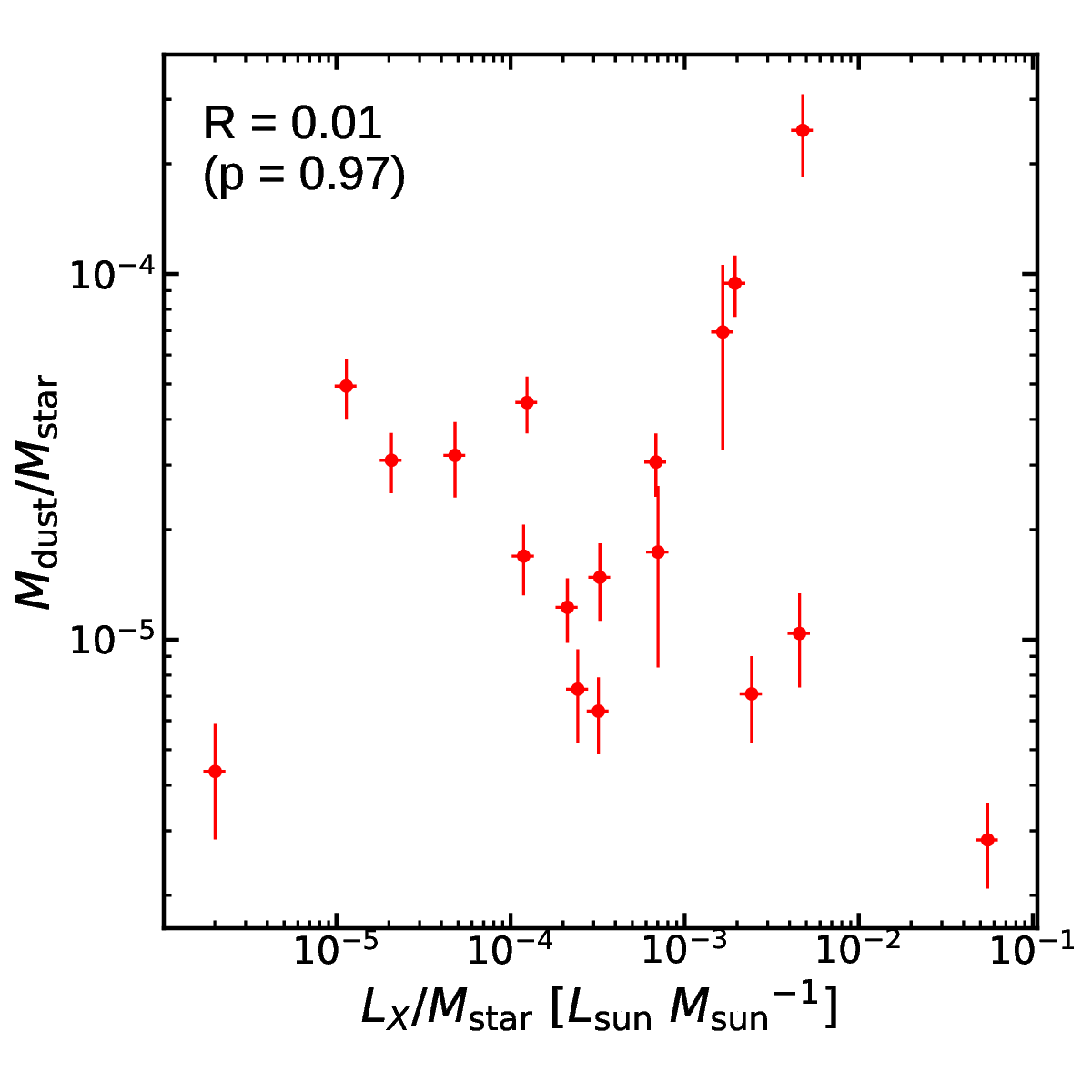} 
 \end{center}
\caption{Scatter plot of $M_\mathrm{dust}/M_\mathrm{star}$ vs $L_\mathrm{X}/M_\mathrm{star}$ for the sample ETGs with available diffuse X-ray measurements. 
{Alt text: One scatter plot.} 
}\label{fig:8}
\end{figure}

\subsection{Properties of the PAHs detected in near-IR spectra}
We find no clear correlation between the PAH emission and the stellar mass loss (figure 6). 
On the other hand, figure 9 shows a positive correlation between the PAH intensity normalized by $M_\mathrm{dust}$ and $L_\mathrm{hot+warm}/M_\mathrm{dust}$. Here, $L_\mathrm{hot+warm}$ denotes the combined luminosity of the hot and warm dust components. This correlation indicates that PAHs are associated with relatively strong UV radiation fields. Such environments may be related to energetic processes associated with AGN activity, including buoyant outflows that transport PAHs from the central regions to the outer parts of ETGs (\cite{Temi2007}; \cite{Kaneda2011}).\par
These results indicate that PAHs in ETGs have an origin different from the cold silicate dust component of the internal origin, and thus likely of an external origin. For example, they may have been supplied together with the gas accreted through galaxy mergers (e.g., \cite{Goudfrooij1994}; \cite{Davis2015}). However, the origin and spatial distribution of the PAHs in the sample ETGs cannot be conclusively determined from the data of this study alone. Spatially resolved spectroscopic observations with high sensitivity, such as those with JWST, are required.
\begin{figure}
 \begin{center}
  \includegraphics[width=8cm]{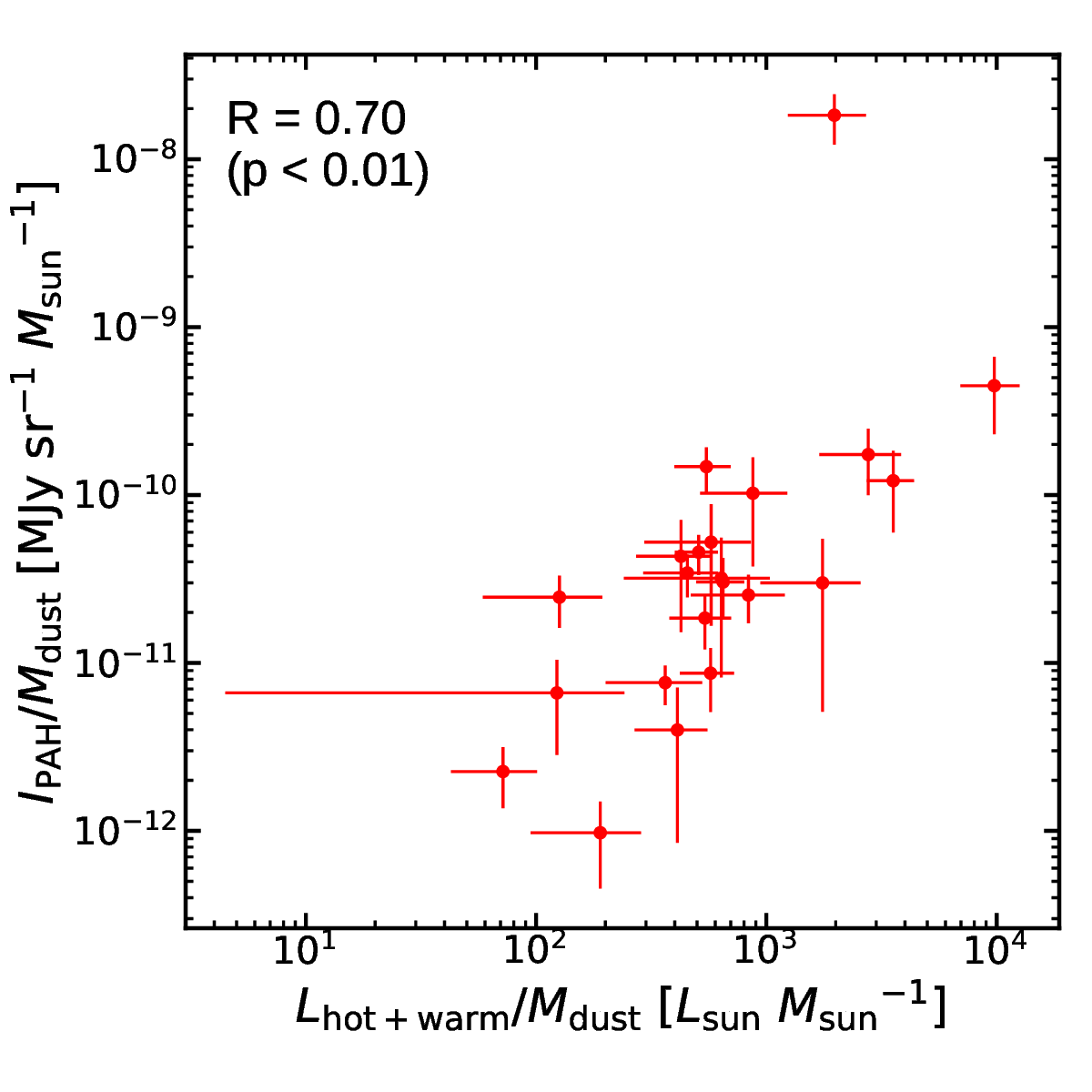} 
 \end{center}
\caption{Scatter plot of $I_\mathrm{PAH}/M_\mathrm{dust}$ vs $L_\mathrm{hot+warm}/M_\mathrm{dust}$ for the sample ETGs detected in the PAH emission. 
{Alt text: One scatter plot.} 
}\label{fig:9}
\end{figure}

\section{Conclusion}\label{sec:5}
We analyzed AKARI near-IR spectra together with mid- to far-IR photometric data for a sample of 30 ETGs to investigate the origin of dust in these galaxies. From the near-IR spectra, we measured the EWs of the SiO and CO absorption features, while the dust masses were estimated by fitting dust emission models to the SEDs constructed from the IR photometry. We find a correlation between the dust mass and the summed EWs of the SiO and CO absorption features, suggesting that stellar mass loss is an important source of dust in ETGs. Since the SiO and CO absorption features trace the photospheres of O-rich AGB stars, this result suggests that the cold dust component in ETGs may be largely composed of internally produced silicate dust. Although the 3.3\>µm PAH feature is detected in the near-IR spectra, its intensity shows no correlation with the EWs of the SiO and CO absorption features, implying that PAHs have a different origin from the silicate dust component.\par
We also examine the effects of dust destruction by sputtering in X-ray plasma. No correlation is found between the dust mass and the X-ray luminosity, and more than half of the sample ETGs contain dust masses exceeding those expected from the balance between stellar mass loss and sputtering destruction. These results indicate that dust destruction by the X-ray plasma is inefficient in ETGs, possibly because dust is embedded in the cold gas phase and thus shielded from the X-ray plasma. We also investigate the origin of the PAHs further to find a correlation between the PAH intensity and the luminosity of the hot and warm dust components, suggesting that the PAHs may be of external origins associated with galaxy merger remnants, heated by the activities of galactic nuclei.

\begin{ack}
This research is based on observations with AKARI, a JAXA project with the participation of ESA. This publication makes use of data products from the Wide-field Infrared Survey Explorer, which is a joint project of the University of California, Los Angels, and the Jet Propulsion Laboratory/California Institute of Technology, funded by the National Aeronautics and Space Administration, and data products from the Infrared Astronomical Satellite (IRAS), which is a joint project of the US, UK, and the Netherlands. This research has made use of the NASA/IPAC Extragalactic Database, which is funded by the National Aeronautics and Space Administration and operated by the California Institute of Technology and use of the SIMBAD database, operated at CDS, Strasbourg, France. This work is financially supported by JST SPRING, Grant Number JPMJSP2125. The principal author is grateful for the “THERS Make New Standards Program for the Next Generation Researchers.”
\end{ack}


\end{document}